\documentclass[11pt]{article}

\usepackage{euscript}
\usepackage{amssymb}
\usepackage{amsfonts}
\usepackage{amsbsy}
\usepackage{amsmath}
\usepackage{epsfig}
\usepackage{amsthm}
\usepackage{amscd}
\usepackage{amstext}
\usepackage{verbatim}

\usepackage[pdftex]{hyperref}

\def\ben{\begin{equation}}
\def\een{\end{equation}}
\def\half{{\textstyle{1\over2}}}

\let\a=\alpha \let\b=\beta  \let\d=\delta 
   \let\k=\kappa
     
\let\s=\sigma

\let\w=\omega

\let\pa=\partial
\def\be{\begin{equation}}
\def\ee{\end{equation}}
\def\ba{\begin{array}}
\def\ea{\end{array}}

\def\dalemb#1#2{{\vbox{\hrule height .#2pt
       \hbox{\vrule width.#2pt height#1pt \kern#1pt
               \vrule width.#2pt}
       \hrule height.#2pt}}}

\newcommand{\bea}{\begin{eqnarray}}
\newcommand{\eea}{\end{eqnarray}}

\def\R{{{\Bbb R}}}

\numberwithin{equation}{section}

\begin{document}

\begin{center}

{ \LARGE {\bf Electron stars for holographic metallic criticality}}

\vspace{1.2cm}

Sean A. Hartnoll and Alireza Tavanfar

\vspace{0.9cm}

{\it Center for the Fundamental Laws of Nature, Harvard University,\\
Cambridge, MA 02138, USA \\}

\vspace{0.5cm}

{\tt hartnoll@physics.harvard.edu, tavanfar@physics.harvard.edu} \\

\vspace{1.3cm}

\end{center}

\begin{abstract}

We refer to the ground state of a gravitating, charged ideal fluid of fermions held at a finite chemical potential as an `electron star'. In a holographic setting, electron stars are candidate gravity duals for strongly interacting finite fermion density systems. We show how electron stars develop an emergent Lifshitz scaling at low energies. This IR scaling region is a consequence of the two way interaction between emergent quantum critical bosonic modes and the finite density of fermions. By integrating from the IR region to an asymptotically AdS$_4$ spacetime,
we compute basic properties of the electron stars, including their electrical conductivity. We emphasize the challenge of connecting UV and IR physics in strongly interacting finite density systems.

\end{abstract}

\pagebreak
\setcounter{page}{1}

\section{The broader context}

A challenge facing contemporary condensed matter theory is the description of a 2+1 dimensional finite density
of fermions interacting with a gapless collective bosonic excitation, such as a spin density wave or emergent gauge field.
Such theories arise, for instance, when a Fermi liquid is tuned across a quantum phase transition. The low energy dynamics of the
system of fermions interacting with the critical bosonic mode can be characterised as metallic quantum criticality. While in 3+1
dimensions one can proceed to integrate out the fermions and obtain a stable Gaussian theory for the boson \cite{hertz}, this approach does not give correct answers in 2+1 dimensions, see e.g. \cite{abanov1, abanov2, Metlitski:2010vm}, as it ignores an infinite number of nonlocal marginal couplings in the effective theory for the boson. One should not integrate out the fermions in this case but rather flow to a scaling regime involving both the boson and fermion fields. The resulting low energy theory is strongly interacting, e.g. \cite{Metlitski:2010vm}.

One might have hoped to perform a (vector) large $N$ analysis as a perturbative handle on the theory. It has recently been demonstrated \cite{sungsik,Metlitski:2010pd,Metlitski:2010vm} that the vector large $N$ expansion breaks down for 2+1 dimensional metallic quantum critical systems. This occurs because a potential IR divergence at high loop order is cured by a self-energy of order $1/N$, leading to extra factors of $N$ in the numerator in certain Feynman graphs. Partially motivated by these difficulties, in this paper we will use the holographic correspondence \cite{Maldacena:1997re, Hartnoll:2009sz, Herzog:2009xv, McGreevy:2009xe} to study a strongly interacting system of gapless bosons with a finite density of fermions. Before proceeding we should note that more traditional approaches to this problem have also been proposed \cite{Nayak:1993uh, Nayak:1994ng, Mross:2010rd} and that our framework does not appear to include ingredients that are likely crucial for applications to the original systems of interest, such as Fermi lines with cold regions as well as hot spots. We will, however, describe the emergence of a low energy scaling regime from the interaction of critical bosons with a finite density of fermions. The essential physics of this process was noted in \cite{Hartnoll:2009ns}.

In the holographic correspondence a charge density is implemented by
a bulk Maxwell field, dual to the current operator in field theory. The asymptotic
boundary value of the Maxwell field determines the chemical potential of the field
theory. This is a UV input the consequences of which we wish to explore at low energies.
In the simplest bulk setup of Einstein-Maxwell theory, the gravitational solution
dual to the finite chemical potential theory is then determined uniquely to be the
planar AdS-Reissner-Nordstrom black hole. This black hole was therefore a natural starting
point for investigations into strongly interacting finite density systems \cite{Son:2006em, Hartnoll:2007ai, Hartnoll:2007ih, Hartnoll:2007ip}. Reviews of this and other earlier work can be found in \cite{Hartnoll:2009sz, Herzog:2009xv, Sachdev:2008ba}.

A conceptually problematic aspect of charged black holes in an applied holography context is their blackness. One is often interested in temperatures much lower than the scale set by the charge density. In this extremal black hole limit the horizon remains present and the actual source of the bulk electrical field remains hidden. Thus, within a bulk effective field theory approach to holography we don't have explicit access to the zero temperature charged degrees of freedom.\footnote{While in some supersymmetric theories we might hope to be able to adiabatically continue the problem to a weakly coupled regime and `count' the degrees of freedom there \cite{Strominger:1996sh}, this does not help us with our objective of understanding the strongly interacting finite density dynamics on its own terms.} Although the universality of a black hole description of charge density is appealing, and may yet have important consequences, it makes it difficult to connect with basic experimental implications of a finite density that depend on the nature of the charge carriers, such as a Fermi surface in the case of fermions. A Fermi surface appears not to be inherent to the gravitational geometry, but depends on the nature of external probes \cite{Lee:2008xf, Liu:2009dm, Cubrovic:2009ye, Faulkner:2009wj, Denef:2009yy}.

It is perhaps fortunate, therefore, that low temperature charged AdS black holes are found to be unstable towards a range of processes that discharge the black hole and can lead to spacetimes without black hole horizons. The instabilities include condensation of charged scalar fields \cite{Gubser:2008px}, Cooper pairing of charged fermions \cite{Hartman:2010fk}, emission of D branes \cite{Yamada:2008em, McInnes:2009zp, Hartnoll:2009ns}, backreaction of a bulk fermionic charge density induced by the local chemical potential \cite{Hartnoll:2009ns}, confinement \cite{Witten:1998zw, Nishioka:2009zj, Horowitz:2010jq}, and perhaps the emergence of underlying lattice degrees of freedom \cite{Sachdev:2010um}. It is not clear at this stage whether all zero temperature charged AdS black holes with a finite size horizon are unstable \cite{Denef:2009tp}. If they are, this fact may be closely tied up with a version of the `weak gravity' conjecture \cite{ArkaniHamed:2006dz}. The instabilities lead to a new zero temperature bulk geometry, often without a finite size horizon (e.g. \cite{Gubser:2009cg, Horowitz:2009ij}). The charge is then carried by explicit bulk fields and we can identify the corresponding field theory operators as responsible for the finite density dynamics.

A fruitful approach taken in previous works in order to explicitly model holographic charge carriers even in the presence of horizons is to add probe D branes into the bulk, see e.g. \cite{Karch:2007pd, Karch:2008fa, Kulaxizi:2008jx, Hartnoll:2009ns}. The limitation of this approach, shared with that of probe fermions \cite{Lee:2008xf, Liu:2009dm, Cubrovic:2009ye, Faulkner:2009wj}, is that it does not capture the two way interaction between the (putatively fermionic) charge carriers and the quantum critical modes.

In this paper we will expand upon section 7.4 of \cite{Hartnoll:2009ns} and describe the electromagnetic and gravitational backreaction of charged fermions on the holographic spacetime geometry. In general this is a very difficult problem as the fermions cannot be treated classically. The coupled fermion-Maxwell-gravity system becomes tractable in a limit in which the fermions may be treated locally in the bulk as an ideal fluid of zero temperature charged free fermions. This approach mirrors the standard description of neutron starts in astrophysics, following the original Oppenheimer-Volkoff-Tolman papers \cite{openn, tolman}. The neutron star equations were generalised to an asymptotically AdS setting in \cite{deBoer:2009wk}. Given that our fermion fluid is charged, we will refer to our solutions as electron stars.

In the following section we set up the equations of motion for a charged ideal fluid in Einstein-Maxwell theory with a negative cosmological constant. Later in section \ref{sec:act} we derive these equations from an action. We discuss the regime of validity of the fluid description depending on the Newton and Maxwell couplings as well as the cosmological constant. In section \ref{sec:sol} we characterise the (planar) electron star solutions to these equations of motion. We show that the IR of the geometry has an emergent Lifshitz scaling and compute the dynamical scaling exponent $z$ as a function of the parameters of the theory. Integrating out from the IR region to the spacetime boundary, we numerically obtain the full electron star solutions and compute their mass and charge. By perturbing the solutions in section \ref{sec:cond} we obtain the electrical conductivity as a function of frequency. We find that the electron star conductivity exhibits a universal low frequency behaviour previously noted in other solutions with an IR Lifshitz scaling.
Our work presents a framework in which the physics of a strongly interacting finite fermion density system can be investigated; we enumerate some of the more pressing open directions in the final discussion section.

\section{Equations of motion: background}
\label{sec:two}

We are interested in 3+1 dimensional zero temperature configurations of a charged perfect fluid in a holographic setting. We will introduce an action principle in a later section, but for the moment will work with equations of motion.
The Einstein-Maxwell equations with a negative cosmological constant and sources are
\be\label{eq:einstein}
R_{ab} - \frac{1}{2} g_{ab} R - \frac{3}{L^2} g_{ab} = \k^2 \left(\frac{1}{e^2}
\left(F_{ac} F_b{}^c - \frac{1}{4} g_{ab} F_{cd} F^{cd} \right) + T_{ab} \right) \,,
\ee
and
\be\label{eq:maxwell}
\nabla_a F^{ba} = e^2 J^b \,.
\ee
Here the perfect fluid energy momentum tensor and current are
\be\label{eq:ff}
T_{ab} = (\rho + p) u_a u_b + p g_{ab} \,, \qquad J_a = \s u_a \,.
\ee
The four velocity $u$ should be normalised so that $u^2 = -1$. The second Bianchi identity requires that the right hand side of (\ref{eq:einstein}) be transverse. Similarly (\ref{eq:maxwell}) requires that the current be conserved. In the above expressions the cosmological constant scale $L$, Maxwell coupling $e$ and Newton constant $\k^2$ are constants while the pressure $p$, energy density $\rho$ and charge density $\sigma$ are fields on spacetime that will be related through the equation of state of the fluid.

For the background we wish to make the following `planar star' ansatz for the metric and Maxwell field
\be\label{eq:metric}
ds^2 = L^2 \left(- f dt^2 + g dr^2 + \frac{1}{r^2} \left( dx^2 + dy^2 \right) \right) \,, \qquad A = \frac{e L}{\k} h dt \,.
\ee
Here $f,g,h$ are functions of the radial coordinate $r$. The pressure and energy and charge densities are also functions of $r$. It is useful to scale out the couplings and write
\be
p = \frac{1}{L^2 \k^2} \hat p \,, \qquad \rho = \frac{1}{L^2 \k^2} \hat \rho \,,\qquad \sigma = \frac{1}{e L^2 \k} \hat \sigma \,.
\ee
The velocity has nonzero component $u^t = 1/(L \sqrt{f})$.

It is straightforward to show that the above Einstein-Maxwell equations are solved provided that the following four equations are satisfied
\bea
\hat p' + \left(\hat p + \hat \rho \right) \frac{f'}{2 f}  - \frac{h' \hat \sigma}{{\sqrt{f}}} & = & 0 \,,  \label{eq:one}\\
\frac{1}{r} \left(\frac{f'}{f} + \frac{g'}{g} + \frac{4}{r}  \right) + (\hat p + \hat \rho) g & = & 0 \,, \\
\frac{f'}{r f} - \frac{h'^2}{2f} + g (3 + \hat p) - \frac{1}{r^2} & = & 0 \,, \\
h'' + \frac{r h'}{2} g \left(\hat p + \hat \rho \right) - g \sqrt{f} \hat \sigma & = & 0 \,. \label{eq:four}
\eea
These are four equations for six variables and so an additional equation of state must be specified in order to close the system. One of the equations is second order. While we could set 
$h' = F $ at this point to obtain purely first order equations in terms of the Maxwell field strength, we are shortly about to include the effects of Thomas-Fermi screening which introduces an explicit dependence on the Maxwell potential $h$.

We will focus in this paper on the case in which the ideal fluid is made from zero temperature charged fermions with mass $m$. Firstly recall that in flat 3+1 dimensional space with chemical potential $\mu$ we would have
\be\label{eq:flat}
\rho = \int_{m}^{\mu} E\, g(E) dE \,, \qquad \sigma = \int_{m}^{\mu} g(E) dE \,, \qquad - p = \rho - \mu \sigma \,. 
\ee
The last of these expressions is the usual thermodynamic relation for the grand canonical ensemble.
We have taken the charge of the fermion to be one in units where the Maxwell action is $\frac{1}{e^2} F^2$. The density of states is
\be
g(E) = \b E \sqrt{E^2 - m^2} \,.
\ee
The constant of proportionality $\beta$ is order one, the exact value is not important for us. We will see shortly that it is a rescaled constant $\hat \b$ that we wish to dial. Finally, if $\mu < m$ then no states above the vacuum are populated and so $\rho = p = \sigma = 0$.

We will work in the approximation in which the fermion physics is determined by the local chemical potential, which is the tangent frame value of the background Maxwell field
\be\label{eq:ansatz}
\mu_\text{loc.} = A_{\hat t} = \frac{A_t}{L \sqrt{f}} = \frac{e}{\k} \frac{h}{\sqrt{f}} \,.
\ee
This `locally flat space' approximation will be shown to be self consistent in an interesting regime of parameters shortly. In section \ref{sec:act} we will derive the relation (\ref{eq:ansatz}) from the same action that implies the ideal fluid-Einstein-Maxwell equations of motion. Substituting into the flat space formulae (\ref{eq:flat}) and scaling the integration variable leads to the `dimensionless' expressions
\be\label{eq:curved}
\hat \rho = \hat \b \int_{\hat m}^{\frac{h}{\sqrt{f}}}  \epsilon^2 \sqrt{\epsilon^2 - \hat m^2} d\epsilon \,, \qquad \hat \sigma = \hat \b \int_{\hat m}^{\frac{h}{\sqrt{f}}}  \epsilon \sqrt{\epsilon^2 - \hat m^2} d\epsilon \,, \qquad - \hat p = \hat \rho - \frac{h}{\sqrt{f}} \hat \sigma \,. 
\ee
Here
\be
\hat \b = \frac{e^4 L^2}{\k^2} \b \,, \qquad \hat m^2 = \frac{\k^2}{e^2} m^2 \,.
\ee
Again, the energy density and other variables vanish if $\frac{h}{\sqrt{f}} < \hat m$. The integrals in (\ref{eq:curved}) are easily performed analytically.
The local free fermion equation of state described by (\ref{eq:curved}) does not include corrections due to gravitational and electromagnetic interactions. We will check below that these corrections are negligible in the regime in which we will work.

The ansatz (\ref{eq:curved}) determines three of our six functions, and therefore we have to check that it is consistent with the four equations (\ref{eq:one}) - (\ref{eq:four}). Indeed this is the case. The first equation (\ref{eq:one}) is in fact closely related to the first law of thermodynamics and is satisfied by (\ref{eq:curved}). The four equations of motion then reduce to the following three equations
\bea
\frac{1}{r} \left(\frac{f'}{f} + \frac{g'}{g} + \frac{4}{r}\right) + \frac{g h \hat \sigma}{\sqrt{f}} & = & 0 \,, \label{eq:a}\\
\frac{f'}{r f} - \frac{h'^2}{2f} + g (3 + \hat p) - \frac{1}{r^2} & = & 0 \,, \\
h'' + \frac{g \hat \sigma}{\sqrt{f}} \left(\frac{r h h'}{2}  - f \right) & = & 0 \,. \label{eq:b}
\eea
In these expressions $\hat p$ and $\hat \sigma$ are given by (\ref{eq:curved}).

Before solving these equations, we should discuss the values of the two free parameters $\hat \b$ and $\hat m^2$. 
Recall that in the classical gravity regime $\k/L \ll 1$.
We will see shortly that the interesting regime we wish to explore in this paper has the `scaled' constant $\hat \b$ of order one. In order to achieve this we therefore need $e^2 \sim \k/L \ll 1$.
Curiously, this is a fairly natural relationship from the point of view of string theory, as it requires
the gravitational (`closed string') coupling to be the square of the Maxwell (`open string') coupling. This usually corresponds to the `probe brane' limit; it is interesting that integrating out fermions charged under a probe brane gauge field results in an order one local backreaction on the spacetime, in the `dimensionless' sense that we mean it.

In the regime in which $\hat \b$ is order one, the dimensionless mass squared is then of order
$\hat m^2 \sim e^2 m^2 L^2$. The operator dual to the bulk fermion might typically have a scaling dimension $\Delta \sim m L \sim e L/\k \sim 1/e \gg 1$, leading to $\hat m^2 \sim 1$. For the moment we will therefore take $\hat m$ to be order one (including $\hat m = 0$).

At this point we can check whether the approximation of using the local flat space results is valid.
One requirement for the flat space treatment is that the density of fermions is large compared to the curvature scale of the geometry. Thus we can compute, under the assumption that $\hat \sigma$ is order one and
that $e^2 \sim \k/L \ll 1$,
\be
\sigma L^3 \sim \frac{L}{e \k} \sim \frac{1}{e^3} \gg 1 \,. 
\ee
Therefore the regime of order one backreaction of the fermions together with the classical gravity limit is compatible with our `Thomas-Fermi' treatment of the fermions. Also compatible with our treatment is the fact that $m L \gg 1$, implying that the Compton wavelength of the fermions is much smaller than the curvature scale (in the massless case one can note that $\mu_\text{loc.} L \gg 1$).

We can also now check the validity of our `mean field' description of gravitational and Maxwell interactions. In this description, the interactions are between local charge and energy densities, but the equation of state determining these densities does not incorporate these interactions. Following \cite{deBoer:2009wk} we can estimate the local effect of interactions through the Boltzmann formula
\be
\frac{d\sigma}{dt} \sim  \sigma^2 v_F \Omega \,.
\ee
Here $v_F$ is the order one Fermi velocity and $\Omega$ is the gravitational or Maxwell total cross section. Using the scaling of various quantities given above, we can easily estimate that the dimensionless quantity $(\sigma \mu_\text{loc.})^{-1} d\sigma/dt \sim e^4 \ll 1$ for both gravitational and Maxwell interactions. Thus the local effect of interactions is parametrically negligible. We now proceed to solve the equations of motion (\ref{eq:a}) - (\ref{eq:b}) treating $\hat \b$ and $\hat m$ as order one free parameters.

\section{Solution to the background equations of motion}
\label{sec:sol}

\subsection{Low energy scaling regime}

\begin{figure}[h]
\begin{center}
\includegraphics[height=180pt]{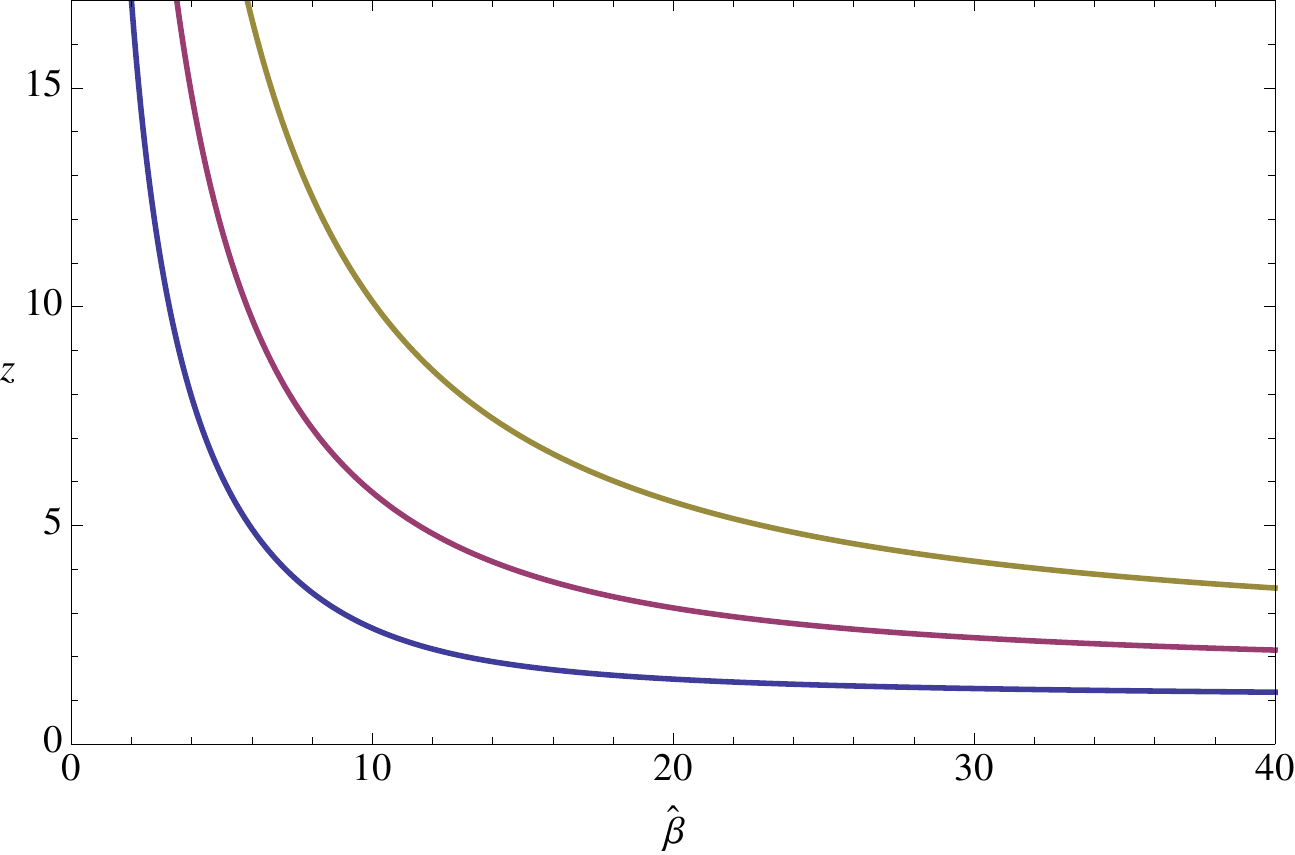}
\end{center}
\caption{Dependence of the IR dynamical critical exponent on $\hat \b$. From left to right, the three curves have $\hat m =0 , 0.55$ and $0.7$.}\label{fig:zbeta}
\end{figure}

In the IR region of the geometry, which will be $r \to \infty$ in our coordinates (\ref{eq:metric}), one finds \cite{Hartnoll:2009ns} an emergent Lifshitz scaling \cite{Kachru:2008yh}. In fact the Lifshitz metric is an exact solution to the equations of motion (\ref{eq:a}) - (\ref{eq:b}). This is perhaps intuitively reasonable: the effect of having a local charge density given by the local background chemical potential (\ref{eq:ansatz}) is to screen the electric field. This might be thought of as a form of Thomas-Fermi screening. Once the electric field has a `mass' it cannot support an $AdS_2$ extremal near horizon geometry. Instead, massive vector fields are known to give rise to Lifshitz solutions \cite{Kachru:2008yh}.
The metric and Maxwell functions take the form
\be\label{eq:lif}
f = \frac{1}{r^{2z}} \,, \qquad g = \frac{g_\infty}{r^2} \,, \qquad h = \frac{h_\infty}{r^z} \,.
\ee
Here $z$ is called the dynamical critical exponent and is given in terms of $\hat \b$ and $\hat m$ by plugging the above Lifshitz ansatz into the equations of motion. From two of the equations of motion we find
\be\label{eq:hg}
h_\infty^2 = \frac{z-1}{z} \,, \qquad g_\infty^2 = \frac{36 (z-1) z^4}{((1-\hat m^2) z - 1)^3 \hat \b^2} \,.
\ee
The remaining equation of motion then gives a complicated relationship between $z, \hat m$ and $\hat \b$ which we cannot solve explicitly.
The dependence of $z$ on $\hat \b$ in plotted in figure \ref{fig:zbeta} below for three values of $\hat m$. Note that the local chemical potential (\ref{eq:ansatz}) is constant on these backgrounds.

It is possible to extract the asymptotic behaviours analytically. For fixed $\hat m$ at large $\hat \b \to \infty$:
\be
z  =  \frac{1}{1 - \hat m^2} +  \frac{6^{4/3} \hat m^{2/3}}{(\hat m^2-1)^{4/3} (2 \hat m^4 - 7 \hat m^2 + 6)^{2/3}} \frac{1}{\hat \b^{2/3}} + \cdots \,. \label{eq:zz}
\ee
In the massless limit the expansion is a little different
\be
z = 1 + \frac{6}{\hat \b} + \cdots \,, \qquad (\hat \b \to \infty \,, \hat m = 0) \,.
\ee
At fixed $\hat m$ and small $\hat \beta \to 0$ (this was the limit considered in \cite{Hartnoll:2009ns}):
\be
z  =  \frac{36}{(1 - \hat m^2)^{3/2}} \frac{1}{\hat \b} - 1 + \frac{3 \hat m^4 \log \frac{1 + \sqrt{1 - \hat m^2}}{\hat m}}{2 (1 - \hat m^2)^{3/2}} + \cdots \,. 
\ee
It is immediately seen that these asymptotic results agree with the behaviour exhibited in the above figure. We see that at intermediate values of $\hat \b$ the dependence of $z$ on $\hat \b$ interpolates between the limiting behavious without any intermediate features.

To make sense of the above expansions, we should first note that if $z \to \infty$ the geometry becomes $AdS_2 \times \R^2$. We see that this occurs as $\hat m \to 1$ from below or as $\hat \b \to 0$ at fixed $\hat m$. In these limits the fermion backreaction is being turned off and one recovers the near horizon geometry of an extremal planar Reissner-Nordstrom-AdS black hole. In particular it is clear that the above solutions only make sense for
\be
0 \leq \hat m < 1 \,.
\ee
For masses bigger than unity, the Lifshitz background chemical potential is not able to induce a density of fermions. It is possible that interesting scaling behaviour arises in the limit $\hat m \to 1$, but we will not investigate this here. From (\ref{eq:zz}) and the above plot we can see that by dialing $\hat \b$ at fixed $\hat m$ we can achieve all $z$ satisfying
\be\label{eq:range}
z \geq \frac{1}{1 - \hat m^2} \geq 1 \,.
\ee

It is also interesting to note that for massless fermions $z \to 1$ as $\hat \b \to \infty$, hence the geometry becomes $AdS_4$. In general, the emergence of an IR Lifshitz scaling geometry with dynamical critical exponent $z$ tunable using couplings and the mass is reminiscent of similar results for holographic superconductors \cite{Gubser:2009cg, Horowitz:2009ij}. The physical difference between the two cases is that for the superconductors the bulk Maxwell field becomes massive due to the Anderson-Higgs mechanism while in the present case the `mass' is due to screening by the charge density. A different type of screening of the Maxwell field, due to a dilaton coupling rather than a charge density, was shown to lead to a Lifshitz IR region in \cite{ Goldstein:2009cv, Taylor:2008tg}.

The IR Lifshitz solution has the dual field theory interpretation of a low energy scaling regime arising from the interaction of a finite density of fermions with emergent critical bosonic modes (the metric and Maxwell fields in the IR of the bulk). To explicitly connect this scaling to the presence of a finite charge density, we need to integrate out to the UV boundary of the spacetime, where the charge density appears as a boundary condition. We do this in the following subsections. The presence of such density-induced emergent quantum criticality is a non trivial and phenomenologically exciting aspect of our models.

\subsection{From the scaling regime to the electron star boundary}

The next step is to flow up the holographic renormalisation group flow. We do this by starting with the Lifshitz IR fixed point of the previous subsection and perturbing it by an irrelevant deformation. We then follow the flow induced by this deformation into the UV by (numerically) solving the differential equations of motion (\ref{eq:a}) - (\ref{eq:b}). Our treatment here is very similar to that of \cite{Gubser:2009cg, Goldstein:2009cv}. The main difference with those works is that the electron star will `end' at some specific radius $r_\text{s}$ where the fluid pressure and charge and energy densities all go to zero.

To perturb away from the scaling solution we can write
\be\label{eq:lif2}
f = \frac{1}{r^{2z}} \left(1 + f_1 r^{\a} + \cdots \right) \,, \qquad g = \frac{g_\infty}{r^2} \left(1 + g_1 r^{\a} + \cdots \right) \,, \qquad h = \frac{h_\infty}{r^z} \left(1 + h_1 r^{\a} + \cdots \right) \,.
\ee
We are looking for solutions where the perturbation grows towards the UV ($r \to 0$) and dies off in the IR ($r \to \infty$). By substituting the above expansion into the equations of motion one easily finds that the three allowed exponents are
\be\label{eq:exponents}
\a_0 = 2 + z \,, \qquad \a_\pm = \frac{2+z}{2} \pm \frac{\sqrt{9 z^3 - 21 z^2 + 40 z - 28 - \hat m^2 z (4- 3 z)^2}}{2 \sqrt{ (1 - \hat m^2) z - 1}} \,.
\ee
As in \cite{Gubser:2009cg, Goldstein:2009cv}, the `universal' relevant deformation with exponent $\a_0$ presumably generates the finite temperature solution, which we will not consider here. The two exponents $\a_\pm$ are both real for the range (\ref{eq:range}) of $z$ and $\hat m$ that we have access to. The exponent $\a_-$ is negative and therefore this is the mode that we need to follow. The $\a_-$ and $\a_+$ modes correspond, respectively, to the coupling and expectation value of an irrelevant operator in the IR theory. At a practical level, the presence of the $\a_0$ and $\a_+$ modes, which must be set to zero for a regular IR Lifshitz region, is why one must numerically integrate from the IR outwards towards the boundary rather than the other way around.

Given the exponent $\a_-$, by series expanding the equations of motion one can determine the coefficients $g_1, h_1$, and all higher coefficients, in terms of $f_1$, which is undetermined. However, $f_1$ can be set to any value by rescaling the coordinates $r,t, \vec x$. This reflects the physical fact that only ratios of dimensionful quantities are meaningful. We can therefore set $f_1$ to an arbitrary constant value (the sign is important however) but should make sure to only compute and plot dimensionless quantities. With the series expansion at hand we can proceed to numerically integrate to smaller values of $r$. A typical result is shown in figure \ref{fig:starplot} below. In the plot we see how the thermodynamic quantities of the fermion fluid flow from their constant Lifshitz values at large $r$ to zero at the star radius $r=r_s$. Note that the IR region of the spacetime, large $r$, has a finite volume in the radial direction.

\begin{figure}[h]
\begin{center}
\includegraphics[height=180pt]{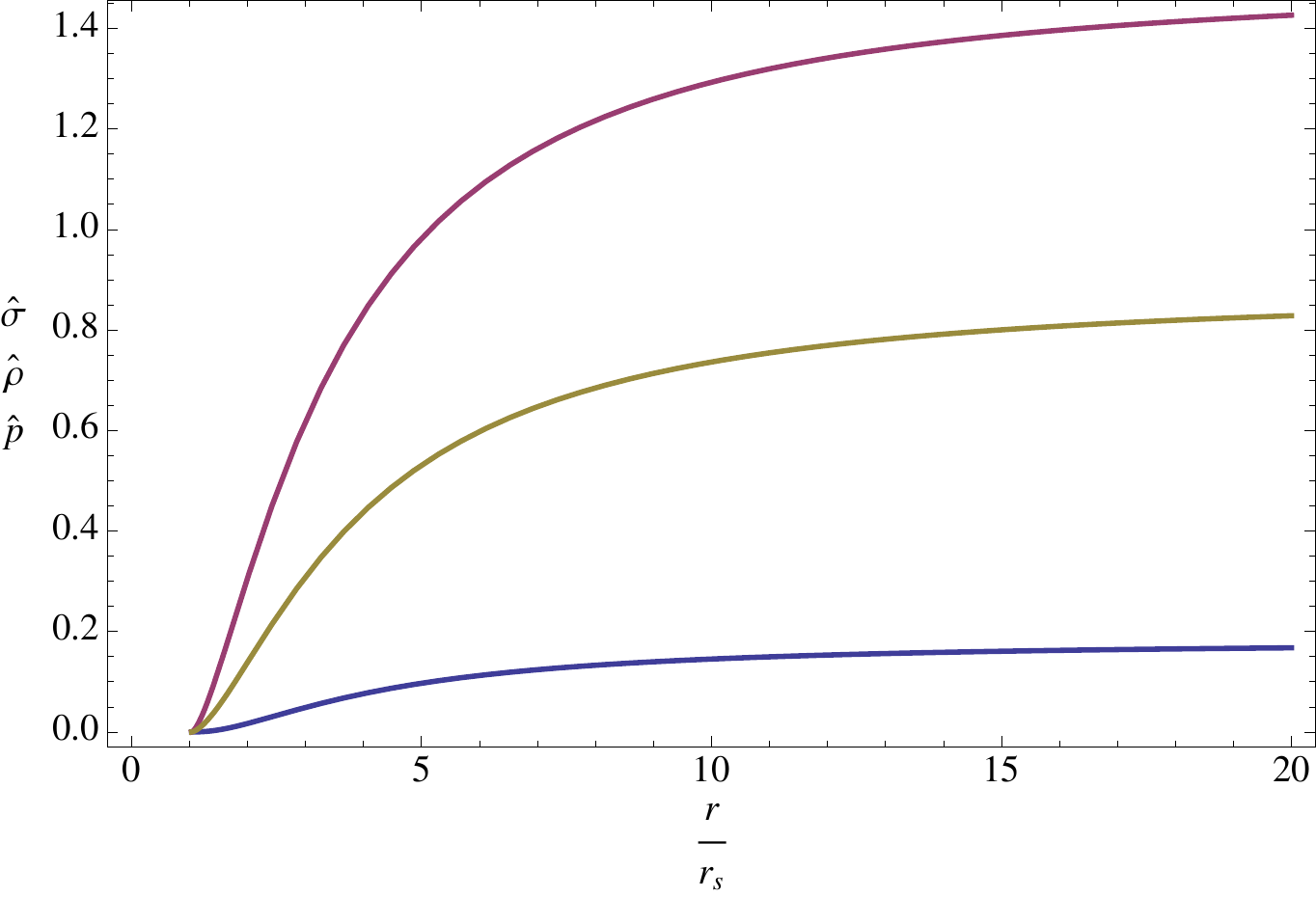}
\end{center}
\caption{From bottom to top, the pressure, energy and charge density distributions for an electron star with $z=2$ and $\hat m = 0.36$ (corresponding to $\hat \b \approx 20$). The boundary of the star is $r=r_s$. Recall that the boundary of spacetime is at $r=0$ while $r \to \infty$ is the deep IR. In the IR the thermodynamic quantities tend to their constant Lifshitz values.}\label{fig:starplot}
\end{figure}

The boundary of the star occurs when the local chemical potential is not large enough to populate the local Fermi sea. Thus from (\ref{eq:curved})
\be
\frac{h(r_s)}{\sqrt{f(r_s)}} = \hat m \,.
\ee

\subsection{Matching onto Reissner-Nordstrom and thermodynamics}

Outside of the electron star $\hat \rho = \hat p = \hat \sigma = 0$ and the solution must become the (planar) Reissner-Nordstrom-AdS spacetime. This solution has
\be\label{eq:RNansatz}
f =  \frac{c^2}{r^2} - \hat M r + \frac{r^2 \hat Q^2}{2} \,, \qquad g = \frac{c^2}{r^4 f} \,, \qquad h = \hat \mu - r \hat Q \,.
\ee
The four constants of integration $\{c, \hat M, \hat Q, \hat \mu\}$, to be related to boundary field theory quantities shortly, must be fixed by matching $\{f, g ,h ,h' \}$ at $r=r_s$. The perhaps unfamiliar constant $c$ is necessary because the normalisation of the time coordinate has been fixed already by our choice of $f_1$ in (\ref{eq:lif2}) in the star interior. We could choose $f_1$ such that $c=1$, but this will not be necessary so long as we consider dimensionless quantities.

It is physically instructive, mimicking the standard astrophysical description of neutron stars, to let $\{c, \hat M, \hat Q\}$ become functions of $r$ and parametrise the solution by (\ref{eq:RNansatz}) throughout the spacetime. It is then a short exercise to show that the functions $\hat M(r)$ and $\hat Q(r)$ obey
\bea
\left(r \hat Q(r)\right)' & = & c(r) \int^{\infty}_r \frac{\sqrt{g(s)}}{s^2} \hat \s(s) \, ds  \,, \\ 
\hat M(r) - \frac{r \hat Q(r)^2}{2} & = & c(r)^2 \int^{\infty}_r \left( \frac{\hat \rho(s)}{s^4} + \frac{h'(s)^2}{2 s^4 f(s) g(s)} \right) ds \,.
\eea
These identities are valid for any ideal fluid and do not depend on the specific equation of state (although we do use the zero temperature thermodynamic relation $-\hat p = \hat \rho - \hat \sigma h/\sqrt{f}$). The above integrals show how the charge $\hat Q$ and energy $\hat M - \half r \hat Q^2$ enclosed within a given radius are determined respectively by the charge density of the ideal fluid and by the sum of the energy density of the fluid and the energy in the electromagnetic field.

By evaluating the previous expressions at the boundary, $r=0$, we obtain formulae for the charge and energy densities of the dual field theory
\bea
\hat Q & \equiv & \hat Q(0) =  c \int^{\infty}_{r_s} \frac{\sqrt{g(s)}}{s^2} \hat \s(s) \, ds \,, \\
\hat E & \equiv & \hat M(0) =  c^2 \int^{\infty}_{r_s} \left( \frac{\hat \rho(s)}{s^4} + \frac{h'(s)^2}{2 s^4 f(s) g(s)} \right) ds + \frac{r_s \hat Q^2}{2} \,. \label{eq:E}
\eea
These quantities are densities with respect to the two boundary spatial dimensions, while $\hat \sigma$ and $\hat \rho$ were densities with respect to the bulk three spatial dimensions. Because the UV theory is a relativistic conformal field theory in 2+1 dimensions, we must have that the pressure and energy are related by $\hat E = 2 \hat P$. Furthermore, in the grand canonical ensemble the free energy $\hat \Omega = - \hat P$. It follows from the zero temperature thermodynamic relation
\be\label{eq:PEQ}
- \hat P = \hat E - \hat \mu \hat Q \,,
\ee
that we must have
\be\label{eq:thermorelation}
\hat E = \frac{2}{3} \hat \mu \hat Q \,.
\ee
In these expressions $\hat \mu$ is the chemical potential of the dual field theory, see e.g. \cite{Hartnoll:2009sz}, not to be confused with the local bulk chemical potential.

The thermodynamic identity (\ref{eq:thermorelation}) can be checked numerically. In deriving this result analytically we were lead to the following  useful observation: on our solutions
\be\label{eq:cool}
2 r h h' - 2 f - r f' = 0 \,.
\ee
We can show this in two steps. Firstly we note that (\ref{eq:cool}) is true on the Lifshitz solution (\ref{eq:lif}) satisfying (\ref{eq:hg}). Secondly, by differentiating the expression in (\ref{eq:cool}) and using the general equations of motion (\ref{eq:one}) - (\ref{eq:four}) together with $-\hat p = \hat \rho - \hat \sigma h/\sqrt{f}$ we can show that it remains zero along radial evolution. Now using (\ref{eq:cool}) it is possible to integrate by parts in (\ref{eq:E}) and derive the identity (\ref{eq:thermorelation}).

First integrals of the equations of motion like (\ref{eq:cool}) are common in gravitational backgrounds and typically implement the isentropy of the classical gravitational flow. In our case we are at zero temperature and the entropy is zero. In practice, we can replace the second order equation (\ref{eq:b}) by the first order relation (\ref{eq:cool}). Thus we have reduced the equations of motion to three first order equations (albeit still involving a first derivative squared).

\begin{figure}[h]
\begin{center}
\includegraphics[height=180pt]{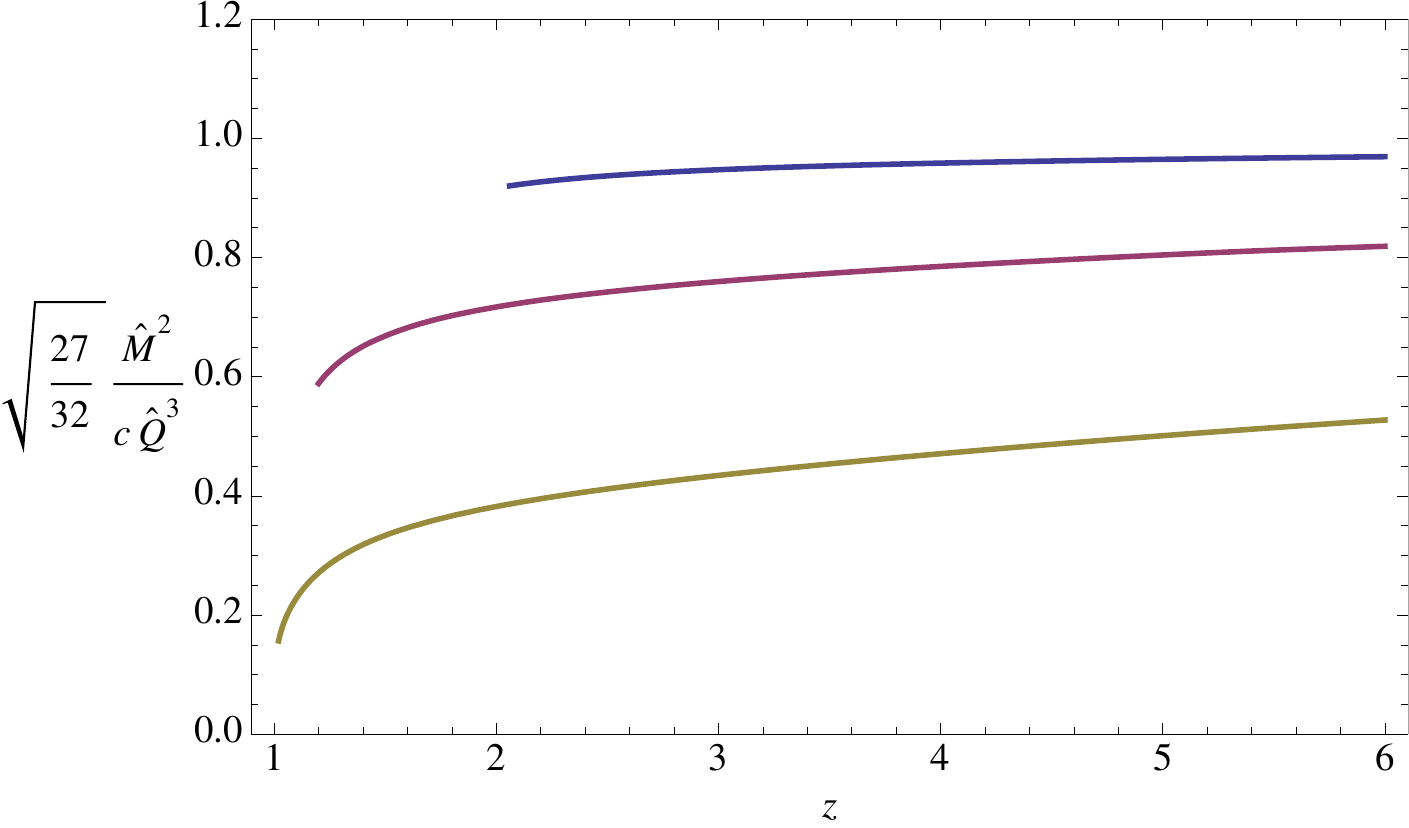}
\end{center}
\caption{Dimensionless ratio of the electron star mass to its charge, normalised such that the ratio is unity for an extremal black hole. The three curves correspond, from top to bottom, to masses $\hat m = 0.7$, $\hat m = 0.36$ and $\hat m = 0.07$.}\label{fig:mqratio}
\end{figure}

To compare the different electron stars at different values of $z$ and $\hat m$, an instructive variable to consider is the dimensionless ratio of the total energy and charge (densities). This ratio can also be compared to the value for extremal Reissner-Nordstrom black holes with no fermionic hair
\be\label{eq:extremal}
\sqrt{\frac{27}{32}} \frac{\hat M^2}{c \hat Q^3} = 1 \qquad (\text{extremal Reissner-Nordstrom}) \,.
\ee
The lower this ratio, the more efficiently the solution is able to carry the charge $\hat Q$. The ratio is shown in figure \ref{fig:mqratio} for various electron stars as a function of the IR critical scaling exponent $z$ for different values of the fermion mass $\hat m$.

In figure \ref{fig:mqratio} we see that the mass of an electron star at fixed charge is always lower than the corresponding extremal black hole with the same charge. The stars are therefore thermodynamically preferred. This might appear surprising as extremal black holes are often the lightest charged objets in the theory with a given charge, being made of `pure charge' in some sense. The situation here is quite analogous to that of holographic superconductors \cite{Hartnoll:2008vx, Hartnoll:2008kx}. Furthermore, we should note that in an extremal black hole background, the local chemical potential $h/\sqrt{f} \to 1$ at the horizon. This can be seen by substituting the extremality condition (\ref{eq:extremal}) into the metric (\ref{eq:RNansatz}). For any fixed fermion mass $\hat m < 1$, this chemical potential becomes greater than the fermion mass before the horizon is reached, and therefore a fermion density is induced. The extremal black hole is thus never a solution to the equations of motion in the range $\hat m < 1$ that we are considering. Alternatively, one could say that the extremal black hole is a solution with an unstable vacuum for the fermion field, in which the fermion states with energies between $m$ and $\mu_\text{loc.}$ are not populated. Consistently with previous remarks, we see that the extremal black hole behaviour emerges as $\hat m \to 1$ or as $z \to \infty$.

\section{An action for charged ideal fluids}
\label{sec:act}

While the equations of motion developed in previous sections are sufficient for many purposes, an action principle often plays a useful role in the holographic correspondence. In this section we pause in our main development to describe an action that recovers all of the equations of motion we have used above.

A simple action for neutral ideal fluids coupled to gravity was formulated by Schutz \cite{Schutz:1970my}. We will start instead from an `off-shell' form of the Schutz action, following \cite{Brown:1992kc} and more closely \cite{Bombelli:1990ze}, coupled to the Maxwell field along the lines of \cite{deRitis:1985uw}. Off-shell refers here to the treatment of a constraint.

For simplicity we begin with the simplest case of `non-rotating' fluids\footnote{Because our fluid is charged, by non-rotating we will mean that $\epsilon^{abcd} u_b \left(\nabla_c u_d + \half F_{cd}/\mu \right) =0$. In practice this condition consistently restricts the degrees of freedom of the fluid to a `potential flow', as will become manifest below.} at zero temperature. The generalization to thermal rotating 
fluids is straightforward, as we will indicate below. An action functional describing non-rotating charged ideal fluids at zero 
temperature, minimally coupled to gravity, is given by
\be\label{act}
S = \int d^4x \sqrt{-g} \left( \mathcal{L}_{{\rm Eins.}} +  \mathcal{L}_{{\rm Mxwl.}} +  \mathcal{L}_{{\rm fluid}} \right) \,,
\ee
where
\be\label{einsmax}
 \mathcal{L}_{{\rm Eins.}} = \frac{1}{2 \kappa^2} \left( R + \frac{6}{L^2} \right) \,, \;\;\;\;\,  \mathcal{L}_{{\rm Mxwl.}} = -\frac{1}{4 e^2} F_{ab} F^{ab} \,,
\ee
and
\be\label{fldlag}
\mathcal{L}_{{\rm fluid}} = - \rho(\s) + \s u^a (\pa_a \phi + A_a) + \lambda (u^a u_a + 1) \,.
\ee
We will see shortly that $u^a$, $\rho$ and $\s$ are the four velocity, the energy density and the charge density of the fluid respectively, 
$\lambda$ is a Lagrange multiplier, and $\phi$ is a  `Clebsch' potential variable associated with the fluid velocity. As previously, we have set the charge of the fermion to be unity and thereby conflated the charge and number densities. Clearly $\phi$ must shift under a gauge transformation in order for the action to be gauge invariant.

We now proceed to derive the equations of motion from (\ref{act}). The variables with respect to which
we vary the action functional are $\lambda, \s, \phi$, the covariant velocity $u_a$, and finally the gauge 
potential $A_a$ and the metric tensor $g_{ab}$.  The variation with respect to the first three of these variables yields the
following equations
\be\label{lam}
\delta \lambda : \;\;\;|u| = -1 \,,
\ee
\be\label{n}
\delta \s : \;\;\;\rho^{\prime}(\s) = u^a (\pa_a \phi + A_a) \,,
\ee
\be\label{phi}
\delta \phi : \;\;\;\nabla_a (\s u^a) = 0 \,.
\ee
Equation (\ref{phi}) is the continuity equation for the fluid current vector
\be\label{eq:current}
J^a \equiv \s u^a \,,
\ee
while (\ref{lam}) is just the 
statement that the fluid four velocity should be timelike. On physical grounds we can identify the left hand side of equation (\ref{n}) as the local chemical potential
\be
\mu (\s) \equiv \rho^{\prime}(\s) \,,
\ee 
and introduce the fluid pressure $p$ through the thermodynamical equation
\be\label{tde}
p(\s) \equiv - \rho(\s) + \s \mu(\s) \,.
\ee
The previous two formulae are simply useful definitions insofar as the equations of motion are concerned.
We can rewrite (\ref{n}) in the form
\be\label{mu}
\mu = u^a (\pa_a \phi + A_a) \,.
\ee
Note that the fluid chemical potential $\mu$ is gauge invariant.

Next comes the varying of the action with respect to $u_b$. This leads to 
\be\label{ua}
\delta u_a : \;\;\; \s (\pa_a \phi + A_a) + 2 \lambda u_a = 0 \,.
\ee
Multiplying the previous equation by $u^a$, the Lagrange multiplier $\lambda$ is determined as
\be\label{lamfix}
\lambda = \frac{\s \mu}{2} = \frac{1}{2}\; (\rho + p) \,.
\ee
Thereby we also obtain
\be\label{eq:normable}
u_a = - \frac{\pa_a \phi + A_a}{\mu} \,.
\ee
This is the so-called `velocity-potential representation', which is implemented directly in `on-shell' variational formulations.

Varying the action with respect to the gauge potential gives the Maxwell equations
\be\label{mxwl}
\nabla_b F^{a b} = e^2 J^a \,,
\ee
where the current was defined in (\ref{eq:current}) above.
Finally the Einstein equations for the geometry,
\be
R_{ab} - \frac{1}{2} g_{a b} R - \frac{3}{L^2} g_{ab} = \frac{\kappa^2}{e^2} \left(F_{ac} F_{b}^{\;c} - \frac{1}{4} g_{ab} F_{cd} F^{cd}\right) +  \kappa^2\; T_{ab}^{{\rm fluid}} \,,
\ee
are obtained upon varying the action (\ref{act}) with respect to $g_{ab}$. The energy-momentum tensor of the fluid is given by
\be
T^{ab}_{{\rm fluid}} \equiv \frac{2}{\sqrt{-g}} \frac{\delta}{\delta g_{ab}}  \int d^4x \sqrt{-g} \;\mathcal{L}_{{\rm fluid}} = 
g^{ab} \mathcal{L}_{{\rm fluid}} - 2 u^{(a} [\s (\partial^{b)} \phi + A^{b)}) +\lambda u^{b)}] \,.
\ee
Using the previous equations of motion, the on-shell energy-momentum is found to take the standard form for an ideal fluid
\be\label{emtif}
T^{ab}_{{\rm fluid}} =  (\rho + p) u^a u^b + p g^{ab} \,.
\ee

We should check that we have indeed recovered all the equations of the previous section from the full action (\ref{act}). The Einstein-Maxwell-fluid equations (\ref{eq:einstein}) -- (\ref{eq:ff}) have been obtained explicitly. The equation of state $\rho(\sigma)$ was defined implicitly through equation (\ref{eq:flat}), by elimination of $\mu$. We can easily check that (\ref{eq:flat}) furthermore implies $\mu = \rho'(\sigma)$ as required. The ansatz we made in section \ref{sec:two} for the metric, Maxwell field and fluid velocity corresponds to setting $\phi=0$ in (\ref{eq:normable}) with the local $\mu$ consequently given by (\ref{eq:ansatz}).\footnote{As usual in holographic setups, we fixed a gauge ambiguity by requiring $A_t$ to vanish at the horizon. Adding a nonzero constant $\pa_t \phi$ in the formula for the gauge invariant local chemical potential (\ref{mu}) would result in a non-Lifshitz invariant IR. A radially dependent $\phi$ would take us outside of our ansatz, introducing radial fluid flow. This choice of requiring $\mu$ to tend to a constant in the IR can presumably be thought of as a choice of fermion vacuum that is regular at the IR `horizon'. Thanks to Tom Hartman for discussions of this point.} Thus we see that all the equations used in previous sections correspond to a solution of the equations of motion following from the action (\ref{act}).

It is now easy to obtain the on-shell form of $\mathcal{L}_{{\rm fluid}}$. This is required for instance to evaluate the free energy of the electron star. Substituting the equations of motion into the action (\ref{fldlag}) gives
\be\label{schutz}
\mathcal{L}_{{\rm fluid}}^{{\rm on-shell}} = p \,.
\ee
Thus the on-shell Lagrangian for the ideal fluid is simply its pressure. Evaluated on our electron star ansatz (\ref{eq:metric}), one can then verify using the equations of motion (\ref{eq:one}) -- (\ref{eq:four}) that the full Lagrangian in (\ref{act}) is a total derivative
\be
\mathcal{L}^{{\rm on-shell}} = \frac{L^2}{\k^2} \frac{d}{dr} \frac{f' - 2 h h'}{2 r^2 \sqrt{f g}} \,.
\ee
The free energy is given by the Euclidean action evaluated on shell. We have just seen that the bulk action becomes a boundary term. In order to obtain a finite answer we must add the boundary Gibbons-Hawking term and intrinsic counterterms. We will not describe this standard process in detail, see e.g. \cite{Hartnoll:2009sz}. The solution near the boundary takes the form (\ref{eq:RNansatz}), this is all that is needed to evaluate the action on shell as there is no contribution from the IR Lifshitz endpoint of the integral.
The upshot is that the free energy density is
\be
\hat \Omega = \hat M - \hat \mu \hat Q \,,
\ee
as we assumed in (\ref{eq:PEQ}) above.

We can also use the action (\ref{act}) as a starting point for Lifshitz holography, with the UV given by (\ref{eq:lif}) rather than $AdS_4$. Lifshitz holography requires additional boundary counterterms. These are most conveniently packaged \cite{Balasubramanian:2009rx,Cheng:2009df} as a series in powers of $|d\phi + A|^2 + \frac{e^2 h_\infty^2}{\k^2}$, as this combination vanishes on the Lifshitz background and it becomes apparent that only a finite number of such terms are necessary.

If, following Schutz \cite{Schutz:1970my}, we wished to obtain the correct equations of motion for the fluid as well as the geometry from the `on-shell' Lagrangian (\ref{schutz}), we would firstly need to bring the equation of state of the fluid $\rho = \rho(\s)$ into the form $p = p(\mu)$. This is achieved through (\ref{tde}) which may be viewed as the usual Legendre transformation.
Subsequently we can take the norm of (\ref{eq:normable}) to express the pressure in terms of the Clebsch potential $\phi$ and the gauge potential $A_a$. This leads to the final Schutz form of the fluid Lagrangian
\be
\mathcal{L}_{{\rm fluid}}^{{\rm Schutz}} = p(\mu) = p(|d \phi + A|) \,.
\ee
This action should be varied with respect to $\phi, A_a$ and $g_{ab}$. This form of the action shows most explicitly that coupling a charged ideal fluid to gravity without rotation or temperature is equivalent to coupling to a St\"uckelberg field.

Finally, we should sketch the straightforward generalization of the off-shell action (\ref{act}) to describe charged rotating and finite temperature ideal fluids coupled to gravity. More details can be found in the papers we referred to above. One first introduces two pairs of new potential-variables, $(\alpha,\beta)$ and $(s,\theta)$. The first pair will account for the fluid rotation.
In the second pair, $s$ will become the fluid entropy density, while the variable $\theta$, the so-called `thermasy', will be responsible for the fluid temperature. Accordingly,
in the action (\ref{act}), the fluid Lagrangian is promoted to the following 
\be\label{fldlag2}
\mathcal{L}_{{\rm fluid}} = - \rho(\s,s) + \s u^a (\pa_a \phi + A_a + \theta \pa_a s + \alpha \pa_a\beta) + \lambda (u^a u_a + 1) \,.
\ee
The equation of state has been enhanced to include an entropy dependence, $\rho(\sigma,s)$.
The equations (\ref{lam}) and (\ref{phi}) will remain the same, while equation (\ref{ua}) is replaced with
\be\label{uag}
\s (\pa_a \phi + A_a + \theta \pa_a s + \alpha \pa_a \beta) + 2 \lambda u_a = 0 \,.
\ee 
There is a similar modification to equation (\ref{mu}).
Further, we have the following equations from varying the action with respect to the new potential variables
\be
u^a \pa_a s = 0\;\;\;;\;\;\; u^a \pa_a \beta = 0\;\;\;;\;\;\; u^a \pa_a \alpha = 0 \;\;\;;\;\;\; u^a \pa_a\theta = - T \equiv - \frac{1}{\sigma} \frac{\pa \rho}{\pa s}\,,
\ee
in which $T$ denotes the fluid temperature. Similarly to above, the resulting Schutz form of the action is found to be
\be
\mathcal{L}_{{\rm fluid}}^{{\rm Schutz}} = p(\mu,s) \,,
\ee
where now the chemical potential in terms of the potential-variables and the gauge field is
\be
\mu = |d \phi + \alpha \, d \beta + \theta \, d s + A| \,.
\ee

\section{Electrical conductivity}
\label{sec:cond}

To compute the frequency dependent electrical conductivity at zero momentum, $\sigma(\w)$, we need to perturb the backgrounds of the previous sections. We clearly need to perturb the vector potential $A_x$, as this is the field dual to the electric current. At zero momentum, these perturbations source perturbations of the metric component $g_{tx}$ and the velocity $u_x$. Specifically, if we take the perturbations to have time dependence $e^{- i \w t}$, so that
\be
A_x = \frac{e L}{\k} \d A_x(r) e^{- i \w t} \,, \qquad g_{tx} = L^2\, \d g_{tx}(r) e^{- i \w t} \,, \qquad u_x = L\, \d u_x(r) e^{- i \w t}  \,,
\ee
then the linearised Einstein-Maxwell equations about the above backgrounds are solved if the following three equations are satisfied
\bea
\hat \sigma \d A_x + (\hat p + \hat \rho) \d u_x & = & 0 \,, \label{eq:fir}\\
\d g_{tx}' + \frac{2}{r} \d g_{tx} + 2 h' \d A_x & = & 0 \,, \\
\d A_x'' + \frac{1}{2} \left(\frac{f'}{f} - \frac{g'}{g} \right) \d A_x' + \frac{h'}{f} \left( \d g_{tx}' + \frac{2}{r} \d g_{tx}  \right)  + g \hat \sigma \d u_x + \w^2 \frac{g}{f} \d A_x & = & 0 \,.
\eea
It is immediately clear that we can eliminate $\d g_{tx}$ and $\d u_x$ from the above equations to obtain a single equation for $\d A_x$
\be\label{eq:ax}
\d A_x'' + \frac{1}{2} \left(\frac{f'}{f} - \frac{g'}{g} \right) \d A_x'  + \left(\frac{\w^2 g}{f}
- \frac{g \hat \sigma^2}{\hat p + \hat \rho} - \frac{2 h'^2}{f} \right) \d A_x= 0 \,.
\ee
The structure of this equation is similar to that arising in holographic superconductors \cite{Hartnoll:2008kx}. There is a `mass' term due to screening in this case rather than electromagnetic symmetry breaking. The rightmost term in the equation (\ref{eq:ax}) is due to the coupling between metric and Maxwell fluctuations. It is also a mass-like term and leads to an infinite DC conductivity because the medium has a net charge density and is translation invariant \cite{Hartnoll:2007ih, Hartnoll:2007ip}.

The equation for the perturbation of the fluid velocity (\ref{eq:fir}) is just $\d A_x + \mu \d u_x = 0$, which is again compatible with the irrotational form (\ref{eq:normable}) with $\d \mu = \d \phi=0$. It is possible that modes with finite momentum $k$ will excite the scalar degree of freedom of the fluid.

\subsection{The conductivity at low frequencies}

In the Lifshitz IR background (\ref{eq:lif}) we can solve the equation (\ref{eq:ax}) for the $\d A_x$ fluctuations analytically in terms of a Hankel function
\be\label{eq:Hankel}
\d A_x^{(\text{Lif.})} = r^{z/2} H^{(1)}_{3/2}\left(g_\infty^{1/2}  \frac{\w r^z}{z} \right ) \,.
\ee
In deriving this formula we imposed ingoing boundary conditions at the Lifshitz `horizon' (i.e. $\d A_x \sim e^{+ i \w g_\infty^{1/2} r^z/z}$ as $r \to \infty$). We also used the algebraic equations (\ref{eq:hg}) giving $g_\infty,h_\infty$ in terms of $z, \hat \b, \hat m$ in order to simplify the index of the Hankel function.
Similar Hankel functions were found for fluctuations of a vector field in a Lifshitz background in e.g. \cite{Goldstein:2009cv, Horowitz:2009ij, Hartnoll:2009ns}. In fact, the use of Hankel functions here is overkill, as the previous expression can equivalently be written as an oscillating exponential multiplying a polynomial
\be\label{eq:LifA}
\d A_x^{(\text{Lif.})} = \left(1 + \frac{i}{\w r^z} \frac{z}{ g_\infty^{1/2}} \right) e^{i  \w r^z\, g_\infty^{1/2}/z } \,.
\ee
We have changed the (unobservable) overall normalisation relative to (\ref{eq:Hankel}).

The solution for the perturbation in the Lifshitz region (\ref{eq:LifA}) will hold for the full solution in the `near' region defined by $r \hat \mu \gg 1$. At zero temperature $\hat \mu$ is the only energy scale in the problem. This condition simply means that we can use the leading order metric near the horizon. At low frequencies, $\w \ll \hat \mu$, the near region has an overlap with the `far' region defined by $\w r^z \hat \mu^{z-1} \ll 1$. In the far region, away from the non-analytic ingoing boundary conditions, we can set $\w=0$ in the Maxwell equation in order to compute to leading order at low frequencies. We will now proceed to match the near solution (\ref{eq:LifA}) to the far region and obtain the conductivity to leading order at low frequencies. This computation is essentially identical to that appearing in e.g. \cite{Gubser:2008wz,Horowitz:2009ij, Goldstein:2009cv}.

The conductivity of the dual field theory is computed at the $AdS_4$ boundary. Near the $AdS_4$ boundary, $r \to 0$, the Maxwell field perturbation behaves as
\be
\d A_x = \d A^{(0)}_x + r \d A^{(1)}_x + \cdots \,.
\ee
The `dimensionless' conductivity is then given by (e.g. \cite{Hartnoll:2009sz})
\be\label{eq:conduct}
\hat \sigma \equiv e^2 \sigma = - \frac{i\, c}{\w} \frac{\d A^{(1)}_x}{\d A^{(0)}_x} \,.
\ee
The extra factor of $c$ compared to more common expressions is due to the normalisation of the metric in \ref{eq:RNansatz}. We use the same symbol for the conductivity of the dual field theory and the charge density in the bulk. Hopefully the context will make it obvious which we are referring to.

The following flux is independent of the radial position $r$
\be
{\mathcal F} = i \sqrt{f/g} \left( \d A_x \overline{\d A_x'} - \overline{\d A_x} \d A_x' \right) \,.
\ee
Constancy of this quantity follows directly from the equation (\ref{eq:ax}). Evaluating near the boundary and using (\ref{eq:conduct}) gives
\be
{\mathcal F} = 2 |\d A_x^{(0)}|^2 \w\, \text{Re} \, \hat \sigma(\w) \,.
\ee
Evaluating using the near horizon solution (\ref{eq:LifA}) gives
\be
{\mathcal F} = 2 \w \,.
\ee
Equating the previous two formulae we see that to obtain the real part of the conductivity it only remains to compute the $\d A_x^{(0)}$ obtained by matching onto (\ref{eq:LifA}). Expanding the near solution (\ref{eq:LifA}) into the matching region $\w r^z \hat \mu^{z-1} \ll 1$ gives
\be
\d A_x  =  \frac{i}{\w r^z} \frac{z}{ g_\infty^{1/2}} \,.
\ee
As we noted above, in the far region we can drop the $\w$ dependence in the Maxwell equation (\ref{eq:ax}) because $\w \ll \hat \mu$. It follows that the $\w$ dependence in the previous formula will remain the same all the way out to the boundary, leading (generically) to $\d A_x^{(0)} \propto \w^{-1}$. Putting all of these facts together then leads to the conclusion that to leading order at low (but finite) frequencies
\be\label{eq:real}
\text{Re}\, \hat \sigma(\w) \propto \w^2 \,.
\ee
This is precisely the same low frequency behaviour for the conductivity in a dual geometry with an IR Lifshitz region as that obtained in \cite{Horowitz:2009ij, Goldstein:2009cv}. The same behaviour also occurs at extremal black hole horizons \cite{Edalati:2009bi}. The physics behind the emergence of the Lifshitz region is apparently distinct in these various cases; this seeming universality in the electrical conductivity remains to be properly understood.

As well as the real part (\ref{eq:real}), we can anticipate that the imaginary part of the conductivity will have a pole as $\w \to 0$ corresponding to a delta function at $\w=0$ in the real part. Thus in fact $\text{Re}\, \hat \sigma(\w) \propto \delta(\w) + \w^2$. This delta function is due to the fact that the system is translationally invariant and carries a net charge. When excited by a time independent electric field, the whole system is accelerated, leading to a current that cannot be relaxed \cite{Hartnoll:2007ih, Hartnoll:2007ip}. We will see the pole in the imaginary part of the conductivity shortly in our numerics.

Before moving on to compute the full conductivity numerically, we should make a remark about the result that $\text{Re}\, \hat \sigma(\w) \propto \delta(\w) + \w^2$. Namely, that there are two reasons why the $\w^2$ conductivity (which will presumably translate into a $T^2$ low temperature dependence of the DC conductivity as in \cite{Goldstein:2010aw}) should not be taken overly seriously as an `experimental' feature of this model. Firstly, in practice, disorder or other physics will smear out the delta function into a Drude peak. The magnitude of this peak could easily dominate, e.g. the temperature dependence of the DC conductivity. Secondly, putting aside the Drude peak, an $\w^2$ or $T^2$ dependence of the conductivity is very weak and corresponds to a huge resistivity, going like $\w^{-2}$ or $T^{-2}$. It will be very easy for any other conduction channel to short-circuit this classical contribution. For instance, one loop processes in the bulk involving charged fermions, analogous to those of \cite{Faulkner:2010da}, will likely have a resistivity that goes to zero at low temperatures or frequencies. The semiclassical expansion will therefore break down at sufficiently low frequencies at which non-classical conduction becomes favoured. It is clearly of interest to investigate the one loop physics of our electron star backgrounds.

\subsection{The full conductivity}

In this subsection we compute the full frequency dependent conductivity. To do this we must numerically solve the differential equation (\ref{eq:ax}). We can integrate out from the horizon to the boundary and then read off the conductivity using (\ref{eq:conduct}).
The ingoing solution at the horizon, $r \to \infty$, on a general background takes the form
\be\label{eq:generalexpand}
\d A_x = e^{i  \w \, g_\infty^{1/2} \left(r^z/z \, + \, r^{z+\a_-}(g_1-f_1)/2 (z+\a_-) \right)} (1 + \# r^{z + 2 \a_-} + \cdots ) \,.
\ee
Here $\a_- < 0$ is the exponent of the IR irrelevant mode in (\ref{eq:exponents}) while $g_1$ and $f_1$ are the coefficients appearing in (\ref{eq:lif2}). These are determined, along with $\#$ and higher order terms, by series expanding the equations of motion.
From the definition of $\a_-$ in (\ref{eq:exponents}) and the lower bound on $z$ in (\ref{eq:range}) we can show that $z + \a_- > 0$ and $z + 2 \a_- < 0$. This is why the first power must be kept in the exponent while the second can be expanded as $r \to \infty$.

Starting from the series expansion (\ref{eq:generalexpand}), we can numerically integrate the Maxwell equation (\ref{eq:ax}) out to the electron star boundary $r_s$. Outside of the electron star, we need to solve the equation in the Reissner-Nordstom-AdS background (\ref{eq:RNansatz}). This must also be done numerically, with the value and derivative of the fluctuation $\d A_x$ matched across the electron star boundary. Integrating out to the boundary then gives the conductivity (\ref{eq:conduct}). We performed these integrations using {\tt NDSolve} in {\textsc{Mathematica}}. The resulting real and imaginary parts of the electrical conductivity as a function of frequency are shown in figure \ref{fig:cond} below.
\begin{figure}[h]
\begin{center}
\includegraphics[height=150pt]{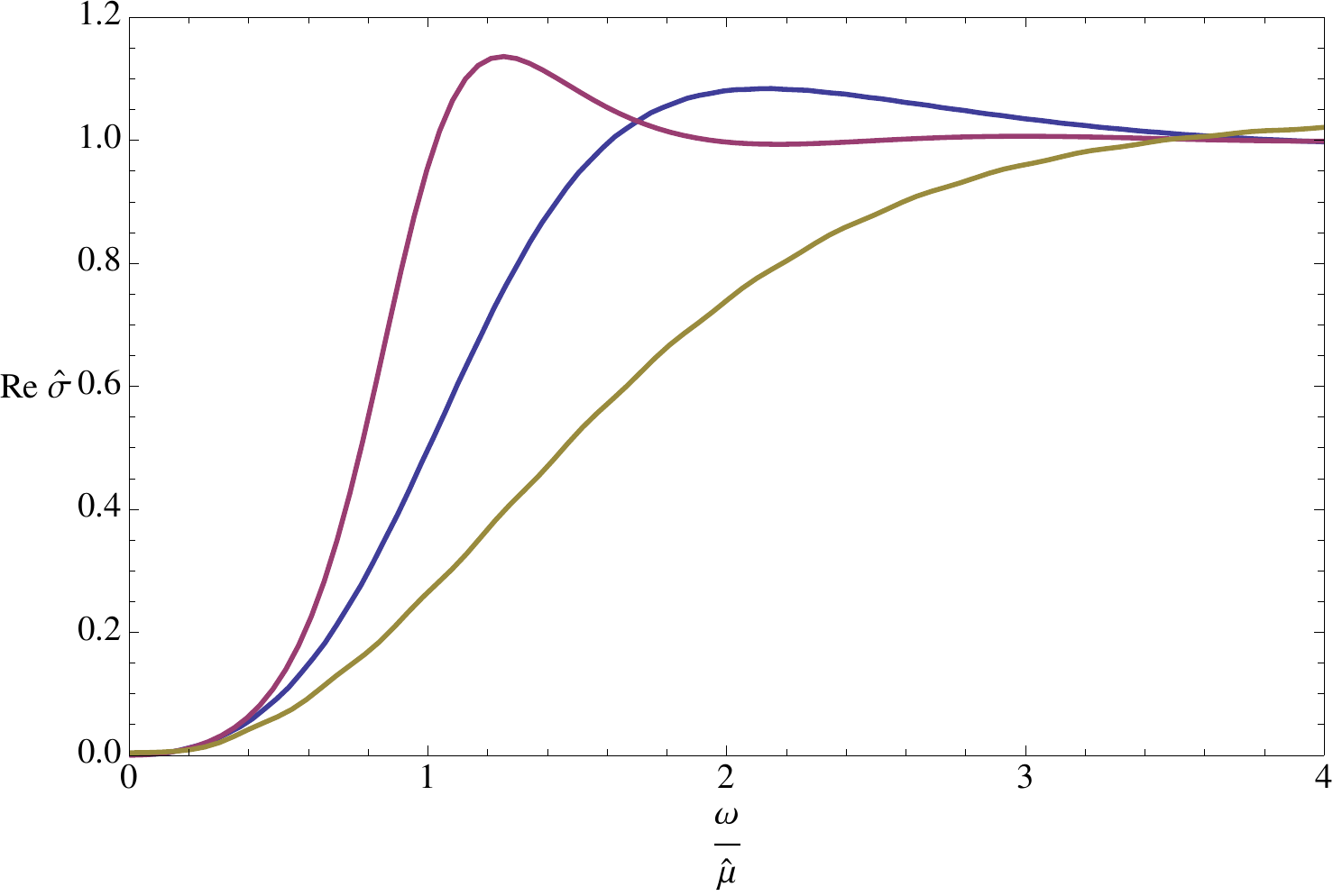}\includegraphics[height=150pt]{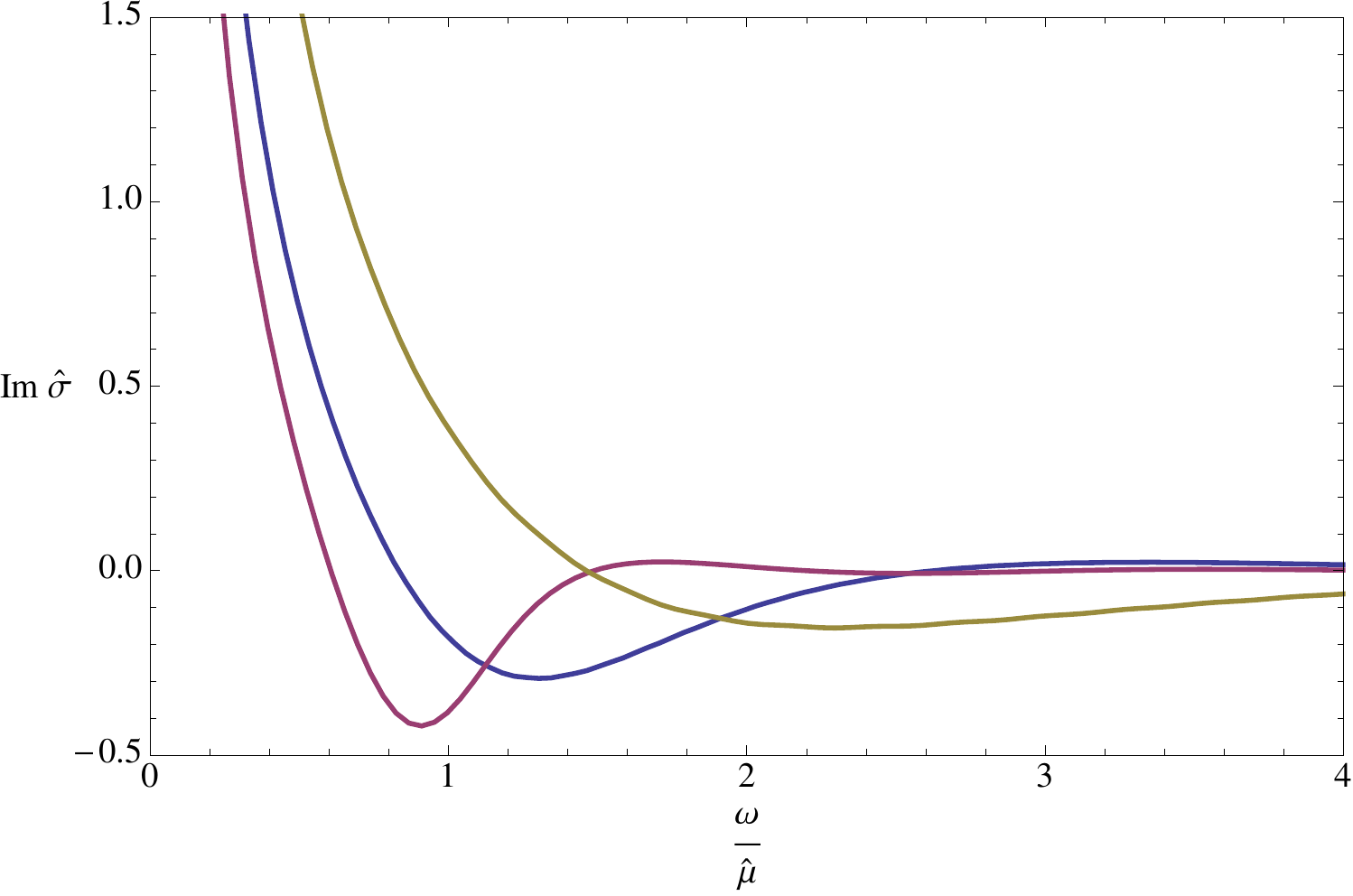}
\end{center}
\caption{The zero temperature real and imaginary parts of the electrical conductivity as a function of frequency. From left to right in each plot: $\{z=3, \hat m = 0.7 \}$, $\{z=2, \hat m = 0.36 \}$ and $\{z=1.5,\hat m = 0.15\}$. The real part also contains a delta function at $\w=0$.}\label{fig:cond}
\end{figure}

In the plots of figure \ref{fig:cond} we see the expected soft gap $\sigma \sim \w^2$ of the (dissipative) real part of the conductivity at low frequencies anticipated in (\ref{eq:real}) above. The divergence of the imaginary part indicates the presence of a delta function in the real part, via for instance the Kramers-Kronig relations. As noted above the divergence is due to the combination of a net charge density and translation invariance (i.e. no impurities or lattice). At large frequencies the real part tends to a constant. This follows from the fact that conductivity is dimensionless in 2+1 dimensions and that the UV completion of our boundary field theory is a conformal fixed point with no inherent scale.

The $z=3$ plot of figure \ref{fig:cond} is already quite similar to the zero temperature limit of the $z=\infty$ Reissner-Nordstrom case, which is a charged black hole rather than an electron star. See e.g. \cite{Hartnoll:2009sz}. The only qualitative effect of the electron star (i.e. lower mass fermions and lower IR scaling $z$) seems to be to smoothen out the transition from the IR $\w^2$ scaling to the constant high frequency behaviour. The coefficient of the $\hat \mu \w^{-1}$ pole in the imaginary part of the conductivity is proportional to $\hat Q/\hat \mu^2 \propto \hat Q^3/\hat E^2$, again see e.g. \cite{Hartnoll:2009sz}. It is therefore consistent with figure \ref{fig:mqratio} that we see that the pole is stronger at lower fermion mass. At lower fermion mass, the electron star has a larger charge at fixed chemical potential.

\section{Final comments}

One important objective of holographic approaches to condensed matter is to characterise possible (computationally controlled) IR fixed point behaviour that falls outside of the Landau Fermi liquid paradigm. As with the Fermi liquid itself, this is in the first instance a question about universal low energy physics, not about the UV physics (be it electrons in a lattice or some cousin of ${\mathcal{N}}=8$ super Yang-Mills theory). This perspective suggests that in the bulk one should focus on the near horizon region of the geometry, as argued most explicitly in \cite{Faulkner:2010tq, Bredberg:2010ky}. However, the role of a finite charge density is subtle in this regard.\footnote{We would like to acknowledge helpful discussions with Diego Hofman on this topic.} The charge density itself, or the chemical potential, is a UV quantity that is specified at the boundary of the bulk geometry. The deep IR Lifshitz solution, in our case for instance, does not immediately `know' what the value of this charge density is. In fact, the electric field is zero at the Lifshitz `horizon' and grows as one moves out towards the boundary of the electron star. Therefore, if we wish to understand how the emergent IR Lifshitz scaling is related to the fact that we are considering a system at finite density, we need to connect statements about the UV and IR physics. In a Fermi liquid, the connection between UV and IR physics is achieved via the Luttinger theorem \cite{lutt1, lutt2}. This theorem states that the volume enclosed by the Fermi surface is determined by the average particle number. When the low energy effective field theory is Fermi liquid theory, the theorem essentially reduces to counting charged states in the UV and IR \cite{oshikawa}. A pressing open question in applications of holography to condensed matter systems is to formulate a useful holographic analogue of this theorem.

Partially motivated to obtain an arena where an interesting low energy scaling geometry could be seen to emerge from a finite density system, in this paper we have constructed zero temperature `electron star' solutions in asymptotically $AdS_4$ spacetime, building on results in \cite{Hartnoll:2009ns}. These geometries are solutions to the Einstein-Maxwell-charged ideal fluid equations of motion. The structure of the solutions is a `domain wall' flow from a `near horizon' Lifshitz geometry to $AdS_4$ at high energies. The essential physics of the flow is that the Maxwell field becomes screened in the Lifshitz region by the charged fluid. In several regards our solutions are qualitatively (and quantitively!) similar to the zero temperature holographic superconductors of \cite{Gubser:2009cg, Horowitz:2009ij} and the extremal dilatonic black holes of \cite{Goldstein:2009cv}.

Another motivation of our work was to obtain gravitational duals in which the full charge density was manifestly `fermionic'. The electron star is literally a Fermi surface that is inhomogeneous in the bulk radial direction. This may or may not be an important ingredient in formulating a holographic Luttinger-like theorem. There are various immediate questions to be explored in this regard. One should compute the momentum dependence of the conductivity to look for Fermi surface related signatures. Also, upon adding a magnetic field to the system the electron star should show quantum oscillations already at a classical level, unlike charged black holes for which quantum oscillations are only present at one loop order and reveal a `small' Fermi surface \cite{Denef:2009yy, Hartnoll:2009kk}.

There are various directions in which our work could be extended at the level of generalising the solutions we have presented and studying their physics. Upon adding interactions the stars will likely have instabilities such as Cooper pairing instabilities along the lines of \cite{Hartman:2010fk}. One should explore the effects of changing the equation of state. There may be circumstances in which clumping instabilities arise. It should be straightforward to place the system at finite temperature. It is possible that a phase transition to a black hole solution occurs at some finite temperature, analogously to the transition in the neutron stars of \cite{deBoer:2009wk}. A more challenging question is to move away from the ideal fluid limit. One would like to solve the Dirac equation in an unspecified background, populate the low lying states up to some chemical potential and then self-consistently solve the Einstein-Maxwell equations together with this quantum source.

\section*{Acknowledgements}

It is a pleasure to acknowledge helpful input from Frederik Denef, Tom Hartman, Chris Herzog, Diego Hofman, Max Metlitski and Subir Sachdev at various points during this work. We are grateful to Tom Hartman, Joe Polchinski and Eva Silverstein for comments on a draft of this paper. Our research is partially supported by DOE grant DE-FG02-91ER40654, NSF grant PHY-0244821 and the FQXi foundation.

\end{document}